# A BAYESIAN-BASED APPROACH FOR PUBLIC SENTIMENT MODELING

Yudi Chen
Wenying Ji

Qi Wang

Department of Civil, Environmental and
Infrastructure Engineering
George Mason University
Fairfax, VA 22030, USA

Department of Civil and Environment
Engineering
Northeastern University
Boston, MA 02115, USA

**ABSTRACT**

Public sentiment is a direct public-centric indicator for the success of effective action planning. Despite its importance, systematic modeling of public sentiment remains untapped in previous studies. This research aims to develop a Bayesian-based approach for quantitative public sentiment modeling, which is capable of incorporating uncertainty and guiding the selection of public sentiment measures. This study comprises three steps: (1) quantifying prior sentiment information and new sentiment observations with Dirichlet distribution and multinomial distribution respectively; (2) deriving the posterior distribution of sentiment probabilities through incorporating the Dirichlet distribution and multinomial distribution via Bayesian inference; and (3) measuring public sentiment through aggregating sampled sets of sentiment probabilities with an application-based measure. A case study on Hurricane Harvey is provided to demonstrate the feasibility and applicability of the proposed approach. The developed approach also has the potential to be generalized to model various types of probability-based measures.

## 1 INTRODUCTION

"Public sentiment is everything," as Abraham Lincoln once said. Government agencies and enterprises have been noticing the importance of public sentiment, and, in turn, employed public sentiment measurement in various applications, such as public opinion mining (O'Connor et al. 2010), customer relationship management (Ang 2011), and enterprise marketing services (Duan et al. 2013), to achieve effective and efficient policy-making and service or product quality improvement. Generally, public sentiment is learned through a sample survey in which a set of people sampled from the interest population are interviewed for their opinions on the issues being considered (Brooker et al. 2003). Although being simple and straightforward, this approach typically solely covers a limited sample size due to labor-intensive and time-consuming interviews (Carter et al. 2001). Recently, the rapid growth and utilization of social media platforms (e.g., Facebook, Twitter, and Instagram), on which people share opinions and sentiments, provide a valuable opportunity for collecting large-scale datasets. However, these large-scale datasets are in fact collections of small datasets with various specific conditions (Ghahramani 2015). For example, in social media-based disaster management, there might be a large amount of disaster-related data over a state, but there is still a relatively small amount of data in each county or city for analyzing local public sentiment and planning relief actions (Ragini et al. 2018). In addition to that, the process of collecting interview- or social media-based sentiments is a random sampling process from the interest population. The smaller the sampled dataset is, the larger uncertainty it includes, and vice versa. Therefore, it becomes necessary to incorporate this inherent uncertainty for accurate and reliable modeling of public sentiment, especially when the sampled dataset is small.

Public sentiment is an aggregation of individuals' sentiments, each of which is typically labeled as negative, neutral, or positive (Beigi et al. 2016). Naturally, sentiment probabilities, which represent ratios for the three sentiments in the interest population, are basic predictors for estimating public sentiment (Yu and Kak 2012). Uncertainty of public sentiment is from the uncertainty of sentiment probabilities that is caused by the statistical variation of the randomly sampled dataset. Recently, researchers have successfully modeled the statistical variation using Beta distribution and binomial distribution to measure binary data uncertainty in the domain of quality control (Ji and Abourizk 2017). This research has inspired us to utilize Dirichlet distribution and multinomial distribution, which are the generalized version of Beta and binomial distributions, to model the uncertainty of sentiment probabilities. Also, the method of aggregating individuals' sentiments heavily depends on applications and requires a systematic review for guiding the selection of an appropriate aggregation method. Therefore, a systematic approach, which is capable of incorporating the inherent uncertainty of the sampled dataset and guiding the selection of the aggregation method, is needed for public sentiment modeling.

The present study aims to develop a Bayesian-based approach to quantitatively model public sentiment in an accurate and reliable manner. Explicitly, the proposed approach (1) incorporates prior sentiment information and sentiment observations via Bayesian inference for enhancing the estimation of public sentiment; (2) guides the selection of an aggregation method for estimating public sentiment from an application-based perspective; and (3) models public sentiment in a systematic and reliable manner through the proposed Bayesian-based approach. The content of this paper is organized as follows. In Section 2, details of the proposed approach are introduced through the steps of sentiment modeling, sentiment integration, and public sentiment measurement. In Section 3, a case study on Hurricane Harvey is presented to demonstrate the feasibility and applicability of the proposed approach. In Section 4, research contributions and limitations are thoroughly discussed.

## 2 METHODOLOGY

As depicted in Figure 1, the proposed approach consists of sentiment modeling, sentiment integration, and public sentiment measurement. First, prior knowledge on sentiment probabilities and evidence from new sentiment observations are modeled through a Dirichlet distribution and a multinomial distribution respectively. Then, the prior knowledge and the new evidence are integrated via Bayesian inference for deriving a posterior distribution on sentiment probabilities. Meanwhile, the posterior distribution is used to update the prior knowledge. Finally, public sentiment is estimated through aggregating sentiment probabilities that are sampled from the posterior distribution.

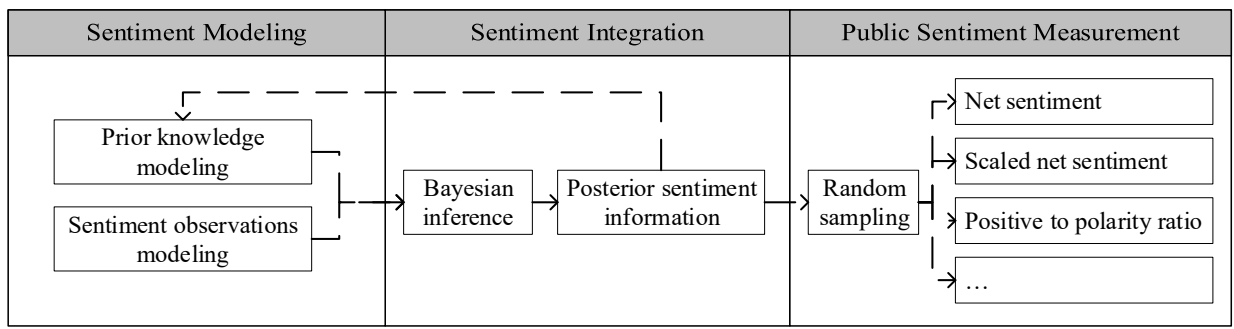

Figure 1: Bayesian-based approach for modeling the public sentiment.

### 2.1 Sentiment Modeling

The modeling of sentiment probabilities, which summarizes and quantifies the sentiment information, is a core step to measure public sentiment. Here, sentiment probabilities are modeled through the integration of prior knowledge and new sentiment observations. In this section, prior knowledge and sentiment observations are modeled with a Dirichlet distribution and a multinomial distribution respectively for deriving the posterior distribution of sentiment probabilities.

### 2.1.1 Prior Knowledge Modeling

In Bayesian inference, a prior distribution of an uncertain variable expresses one's belief on this variable before new evidence is taken into account. In this research, variables in interest are sentiment probabilities for which a Dirichlet distribution is chosen as the prior due to following reasons: (1) sentiment probabilities are bounded within the range of [0, 1] which matches the natural boundaries of a Dirichlet distribution; (2) the property that the Dirichlet distribution is a conjugate prior to the likelihood function of new sentiment observations provides conveniences for deriving an analytical posterior distribution; and (3) parameters of the Dirichlet distribution are intuitively and physically meaningful and easy to estimate from sentiment observations. Hereafter, we use exact numbers {1, 2, 3} to represent the sentiments {negative, neutral, positive} for making following formulas concise and descent. Therefore, prior knowledge on sentiment probabilities $\theta_1, \theta_2, \theta_3$ is modeled as

$$p(\boldsymbol{\theta}) = p(\theta_1, \theta_2, \theta_3) = Dir(\alpha_1, \alpha_2, \alpha_3) = \frac{1}{B(\alpha_1, \alpha_2, \alpha_3)} \prod_{i=1}^{3} \theta_i^{\alpha_i - 1}.$$

where $B$ is a multivariate Beta function for normalization, $\alpha_1, \alpha_2, \alpha_3$ are three shape parameters that control the shape of a Dirichlet distribution. Figure 2 depicts four Dirichlet distributions with various combinations of shape parameters. Color blue represents low probability density, while color red represents high probability density. All sets of sentiment probabilities are constrained over a simplex in $\mathbb{R}^3$.

Ideally, a prior distribution is constructed to match an expert's belief on the variables. However, such belief is a mental condition and heavily depends on the expert's knowledge to a specific condition(Tokdar 2013). The determination of a prior distribution, especially for small datasets, is beyond the scope of this paper in spite of the importance. In this research, a uniform (i.e., non-informative) prior distribution, Dir(1, 1, 1), is used for demonstrating the approach.

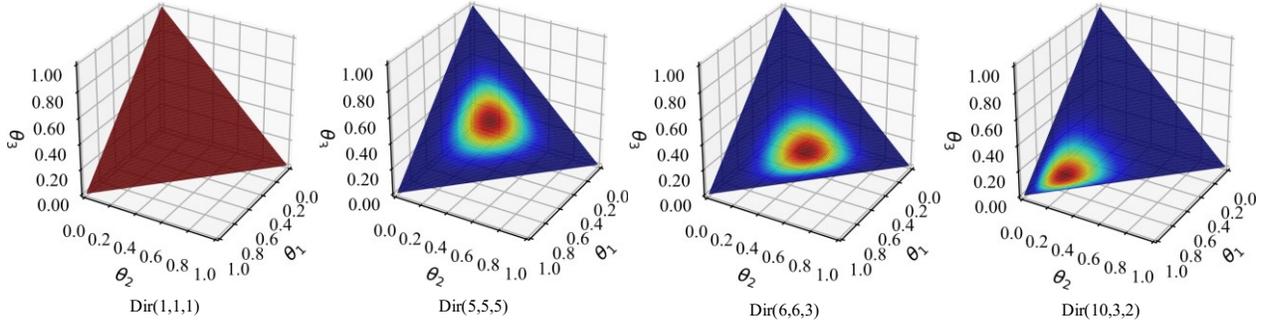

Figure 2: Probability density plots for Dirichlet distributions with various sets of parameters.

### 2.1.2 Sentiment Observations Modeling

In Bayesian inference, new evidence or observed data is generally modeled as a likelihood function for updating prior information. Here, the likelihood function is a multinomial distribution since sentiment is a categorical variable with three identified polarities (i.e., negative, neutral, and positive). Suppose that there are $n$ sentiment observations in the sampled dataset where each observation produces exactly one of the sentiments. Out of $n$ sentiment observations, $x_1$ observations are with negative sentiment, $x_2$ observations are with neutral sentiment, and $x_3$ observations are with positive sentiment. Therefore, the likelihood function is expressed as

$$L(\boldsymbol{X}|\boldsymbol{\theta}) = L(x_1, x_2, x_3|\theta_1, \theta_2, \theta_3) = \frac{\Gamma(\sum_i x_i + 1)}{\prod_i \Gamma(x_i + 1)} \prod_i \theta_i^{x_i}.$$

where $\Gamma$ is a Gamma function.

## 2.2 Sentiment Integration Via Bayesian Inference

Bayesian inference is a method of statistical inference in which Bayes' theorem is used to update the probability for parameters or variables as more evidence or information becomes available (Gelman et al. 2013). Bayesian inference derives the posterior distribution through the integration of two antecedents, the prior $p(\boldsymbol{\theta})$ and the likelihood function $L(\boldsymbol{X}|\boldsymbol{\theta})$, which represent the prior knowledge and new sentiment observations respectively. As described in Section 2.1, the prior knowledge is modeled with a Dirichlet distribution that is a conjugate prior to the likelihood function. Therefore, the posterior distribution is another Dirichlet distribution and is expressed as

$$p(\boldsymbol{\theta}|\boldsymbol{X}) = \frac{L(\boldsymbol{X}|\boldsymbol{\theta})p(\boldsymbol{\theta})}{p(\boldsymbol{X})} = Dir(\alpha_1 + x_1, \alpha_2 + x_2, \alpha_3 + x_3).$$

Notice that, the three shape parameters $\alpha_1 + x_1$, $\alpha_2 + x_2$, and $\alpha_3 + x_3$ of the posterior are simple summations of the corresponding shape parameters $\alpha_1, \alpha_2, \alpha_3$ in the prior distribution and the numbers of observations $x_1, x_2, x_3$ for the three types of sentiments in the sampled dataset. This intuitive property provides an easy and efficient way for deriving the posterior distribution to update beliefs of sentiment probabilities. It also makes the shape parameters of the posterior distribution physically meaningful.

## 2.3 Public Sentiment Measurement

Practically, public sentiment is measured through the aggregation of sentiment probabilities that summarize the sentiment information in the interest population (Yu and Kak 2012). A public sentiment measure—the way of aggregating sentiment probabilities—is determined by the presented sentiment information associated with a specific application. For example, in a movie sales prediction, positive to negative ratio is used as the public sentiment measure to track sentiment variations in different stages (i.e., before and after the movie release) (Zhang and Skiena 2009). Table 1 lists the most commonly used public sentiment measures and gives a brief summary for each of them, including their formulas, boundaries, and applications. Other than these listed measures, public sentiment measure can also be designed to accommodate the specific sentiment information according to the application.

Table 1: Summary of the commonly used public sentiment measures.

| Measure | Formula | Boundary | Presented sentiment information |
|---|---|---|---|
| Net sentiment (Makrehchi 2013) | $\theta_3 - \theta_1$ | $[-1, 1]$ | Sentiment differences for obtaining an overall score |
| Scaled net sentiment (Lică and Tută 2011) | $100 \times (\frac{\theta_3 - \theta_1}{2} + 0.5)$ | $[0, 100]$ | Sentiment differences in a more interpretable range |
| Positive sentiment probability (Zhang and Skiena 2009) | $\theta_3$ | $[0, 1]$ | Positive sentiments |
| Negative sentiment probability (Zhang and Skiena 2009) | $\theta_1$ | $[0, 1]$ | Negative sentiments |

| Positive to polarity ratio (Nguyen et al. 2012) | $\theta_3 / (\theta_1 + \theta_3)$ | [0, 1] | Sentiment change instead of absolute sentiment values |
| --- | --- | --- | --- |
| Positive to negative ratio (Asur and Huberman 2010) | $\theta_3/\theta_1$ | [0, ∞] | Sentiment change with a larger variance |

To have an intuitive understanding of the listed public sentiment measures, histograms of public sentiment based on these measures are shown in Figure 3 with a given posterior distribution on sentiment probabilities, Dir(4, 15, 3). This posterior distribution is consistent with most real applications where data with neutral sentiment occupies a large portion in the dataset (Nakov et al. 2016). For the net sentiment and scaled net sentiment, the distribution shapes are almost same and tend to be normal but with different sentiment ranges. Both the positive sentiment probability and negative sentiment probability are right-skewed, and they are actually Beta distributions since the remaining two probabilities (neutral+negative or neutral+positive) can be taken as one probability, non-positive or non-negative. Compared with the positive to polarity ratio, the positive to negative ratio has no upper limit and is possible to have extreme values.

In the following case study, the negative sentiment probability (NSP) measure will be used for estimating public sentiment for enhanced disaster management due to its representativeness to public demands (Ragini et al. 2018).

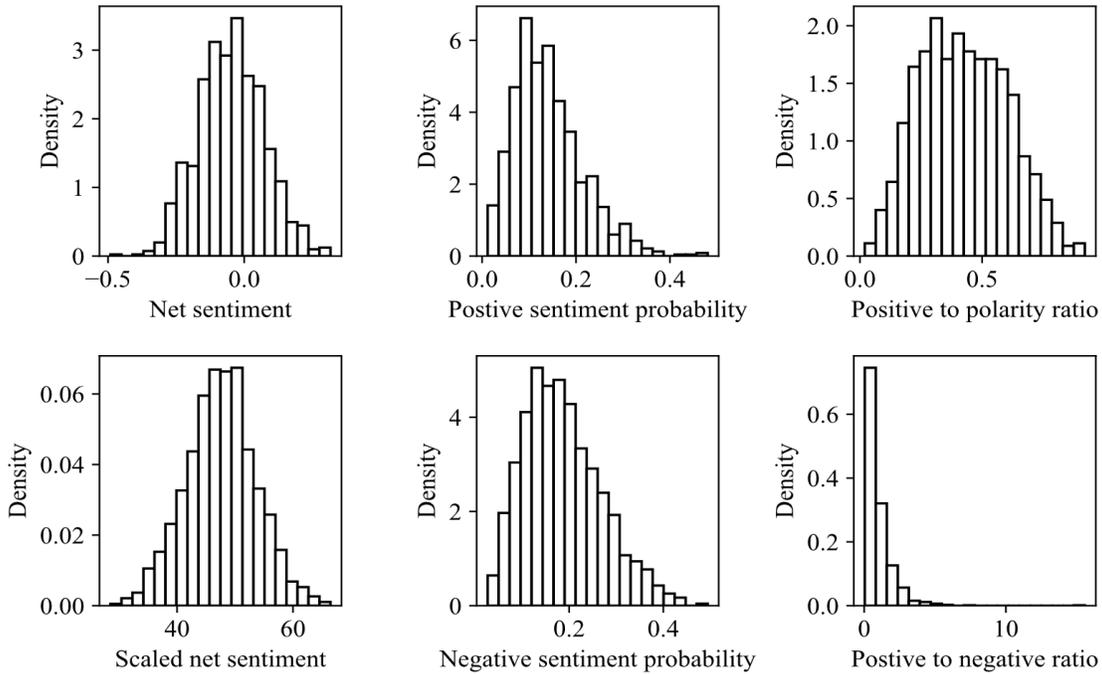

Figure 3: Histograms of public sentiment based on various measures.

## 3  CASE STUDY

In the era of information age, social media platforms are playing an increasingly central role in measuring public sentiments (O'Connor et al. 2010). Twitter, with its social and informational characteristics, is well suited for a fast-paced emergency environment in collecting nearly real-time public sentiment information (Kryvasheyeu et al. 2016). In this section, Tweets, which are posted in Houston during Hurricane Harvey, tied with Hurricane Katrina as the costliest tropical cyclone on record (National Hurricane Center 2018), are collected to model public sentiment through the proposed Bayesian-based approach.

## 3.1 Data Description

The Twitter streaming API was used for collecting geotagged tweets (Wang and Taylor 2016). The collected geotagged tweets were filtered to ensure that the modeled public sentiment is related to Hurricane Harvey and localized in Houston, TX. First, we re-extracted Harvey-related tweets posted during Hurricane Harvey (i.e., Aug. 24, 2017 to Sep. 21, 2017) which contain one of the Harvey-specific keywords (i.e., Hurricane, Harvey, and HurricaneHarvey). Then, the Google map API was used to assign the location information (e.g., country, state, and city) to the Harvey-related tweets, and tweets posted in Houston, TX were further extracted. Totally, 2,963 tweets were used for modeling Harvey-related public sentiment.

Hurricane Harvey brought catastrophic rainfall-triggered flooding in the Houston metropolitan area (Blake and Zelinsky 2018). Figure 4 depicts the daily number of tweets (represented by color blue) and the precipitation in Houston (represented in color red) obtained from the National Oceanic and Atmospheric Administration (NOAA). Similar trends are observed in numbers of daily tweets and precipitation. The numbers of daily tweets first increased dramatically during Aug. 24, 2017 to Aug. 27, 2017 due to the extensive rainfall and reached the peak on Aug. 28, 2017. Unlike the sharp decline in precipitation, the numbers of daily tweets gradually fall off due to the slower process of flood receding and post-disaster recovery.

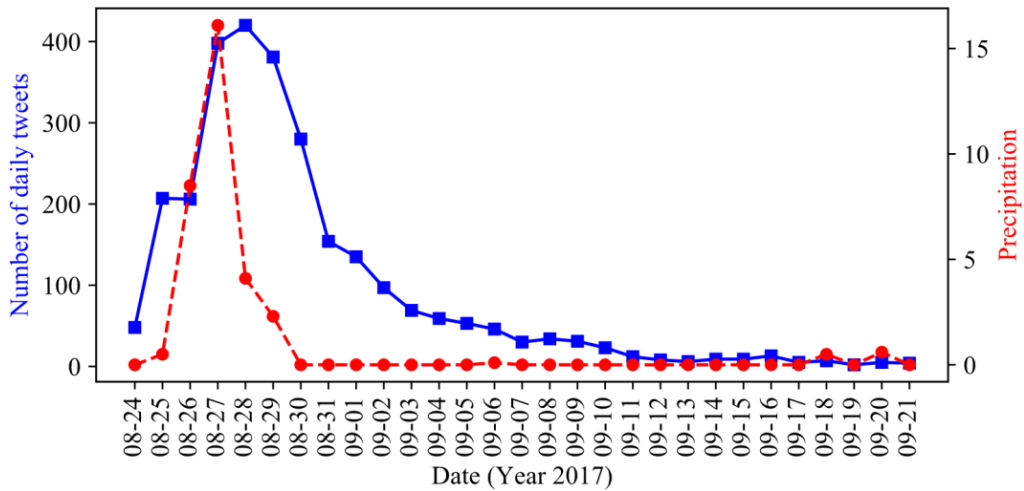

Figure 4: Number of daily tweets and precipitation in Houston during Hurricane Harvey.

## 3.2 Sentiment Prediction

In this section, sentiment for each extracted tweet is predicted as negative, neutral, or positive using SentiStrength, a lexicon-based approach (Thelwall 2017). The core of SentiStrength is a lexicon of 2310 sentiment words and word stems obtained from the Linguistic Inquiry and Word Count (LIWC) program (Pennebaker et al. 2003), the General Inquirer list of sentiment terms (Stone et al. 1966) and ad-hoc additions made during testing. In addition to the lexicon, SentiStrength includes a list of emoticons and a list of idioms together with human-assigned sentiment scores for making accurate sentiment predictions, especially in social media data where emoticons and idioms are extensively used. In most tested social media cases, SentiStrength works consistently well and approaches human-level accuracy (Thelwall 2017). We, therefore, applied the Java-version SentiStrength in this case study. For illustration purposes, Figure 5 shows three sample tweets with negative, neutral and positive sentiments respectively.

As depicted in Figure 4, most tweets and precipitation are observed in the first two weeks (Aug. 24, 2017 to Sep. 6, 2017). Therefore, for illustration purpose, only tweets posted in the first two weeks are analyzed for public sentiment modeling. Figure 6 is a summary of the numbers of daily observations for each sentiment, color blue represents negative, color gray represents neutral, and color red represents positive. These results indicate that tweets with neutral sentiment dominate the data set mainly due to that a large portion of the extracted tweets is for information sharing purposes (e.g., disaster situation report and relief resources sharing) instead of public opinion expression.

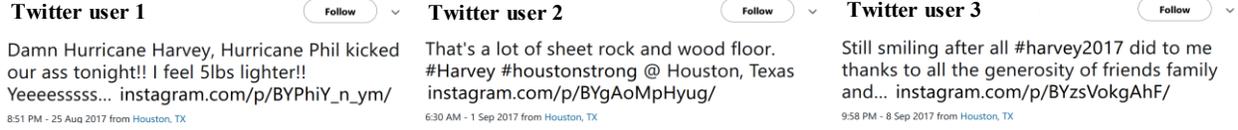

(a) Negative sentiment  (b) Neutral sentiment  (c) Positive sentiment

Figure 5: Sample tweets with negative, neutral and positive sentiment respectively.

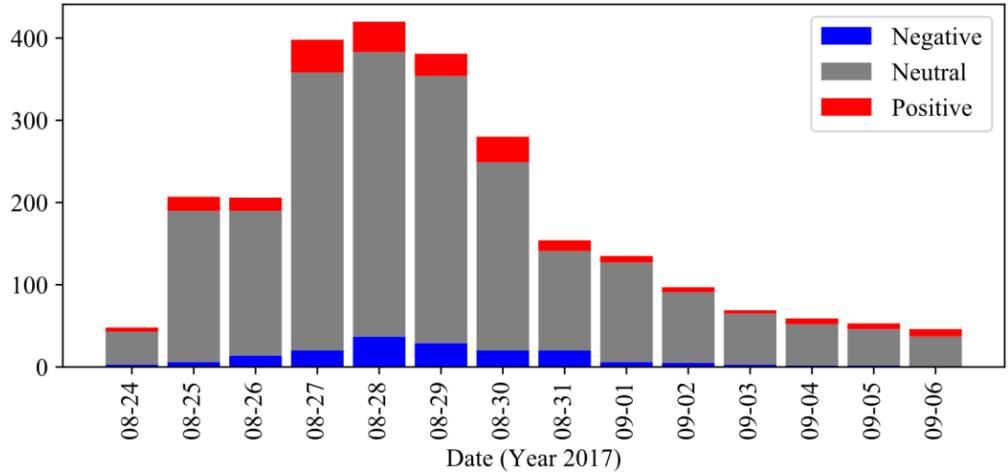

Figure 6: Numbers of daily tweets for each sentiment polarity.

### 3.3 Public Sentiment Modeling

As described in previous sections, public sentiment is measured through aggregating sentiment probabilities, which summarize the sentiment information in the interest population, with a public sentiment measure. Here, negative sentiment probability (NSP) is employed to measure public sentiment since negative sentiment generally represents public demands (Ragini et al. 2018). The sentiment probabilities are sampled from the posterior distribution which is derived by integrating the sentiment information in prior knowledge and daily tweets via Bayesian inference. Table 2 lists the utilized sentiment information, including the prior, numbers of daily tweets for each sentiment, and the posterior distribution of sentiment probabilities. To estimate the daily NSP, the following steps are conducted. First, 1,000 sets of sentiment probabilities $(\theta_1, \theta_2, \theta_3)$ are randomly sampled from the corresponding daily posterior distribution. And then, the NSP $\theta_1$ in each set is saved for constructing the distribution of daily NSP. To visually represent and compare the daily NSP, a side-by-side plot in terms of the mean and 95% credible interval is generated, as shown in Figure 7. The red square is the mean value of daily NSP, and the blue bars are the lower and upper limit of the credible interval. The credible interval is inversely proportional to the uncertainty of daily NSP and a smaller credible interval indicates that the daily NSP is measured with a larger daily tweet dataset. The NSP ranges from nearly 0 to 0.12 and this is because tweets with neutral sentiment dominate in daily tweets. The mean value of the daily NSP first increases rapidly and reaches the peak at Aug. 28, then gradually declines. In addition, the uncertainties during Aug. 27 to Aug. 29 are much smaller than other dates due to the large size of available daily tweets. These observed patterns are intuitively understandable. In the beginning phase of Hurricane Harvey (Aug. 24, 2017 to Aug. 28, 2017), negative sentiment surges due to extensive inconveniences and severe damages caused by the extreme rainfall brought by Hurricane Harvey, and then progressively declines due to the ongoing relief actions in the following recovery phase. These consistencies between the observed patterns and intuitive understanding of human reactions to disasters verify the applicability and feasibility of the proposed modeling approach.

Table 2: Summary of the sentiment information.

| Date (Year 2017) | Prior Distribution | Numbers of daily tweets for each sentiment $(x_1, x_2, x_3)$ | Posterior Distribution |
|---|---|---|---|
| Aug. 24 | Dir(1, 1, 1) | (2, 41, 6) | Dir(3, 42, 7) |
| Aug. 25 | Dir(1, 1, 1) | (7, 185, 18) | Dir(8, 186, 19) |
| Aug. 26 | Dir(1, 1, 1) | (15, 177, 17) | Dir(16, 178, 18) |
| Aug. 27 | Dir(1, 1, 1) | (31, 339, 41) | Dir(32, 340, 42) |
| Aug. 28 | Dir(1, 1, 1) | (38, 347, 38) | Dir(39, 348, 39) |
| Aug. 29 | Dir(1, 1, 1) | (30, 326, 28) | Dir(31, 327, 29) |
| Aug. 30 | Dir(1, 1, 1) | (21, 230, 32) | Dir(22, 231, 33) |
| Aug. 31 | Dir(1, 1, 1) | (11, 122, 14) | Dir(12, 123, 15) |
| Sep. 01 | Dir(1, 1, 1) | (7, 122, 9) | Dir(8, 123, 10) |
| Sep. 02 | Dir(1, 1, 1) | (6, 87, 7) | Dir(7, 88, 8) |
| Sep. 03 | Dir(1, 1, 1) | (4, 63, 5) | Dir(5, 64, 6) |
| Sep. 04 | Dir(1, 1, 1) | (3, 51, 8) | Dir(4, 52, 9) |
| Sep. 05 | Dir(1, 1, 1) | (3, 45, 8) | Dir(4, 46, 9) |
| Sep. 06 | Dir(1, 1, 1) | (1, 38, 10) | Dir(2, 39, 11) |

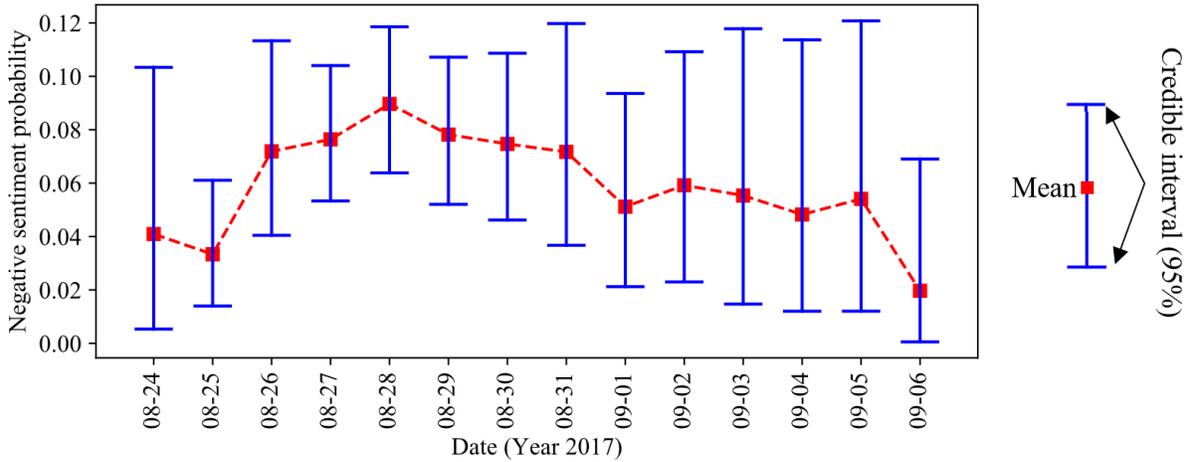

Figure 7: Daily negative sentiment probability for the first two weeks.

## 4 CONCLUSION

Public sentiment is a critical yet often uncertain factor that is highly correlated with the success of action planning in various applications. In this research, a novel Bayesian-based approach is inventively proposed to model public sentiment to integrate the sentiment information from the prior knowledge and sentiment observations in a systematic and reliable manner. In addition, a systematic guidance for the selection of a public sentiment measure is further discussed. A real case of modeling Hurricane Harvey-related public sentiment in Houston, TX is used as an example to demonstrate the feasibility and applicability of the proposed approach.

This research contributes to the academia by (1) developing a Bayesian-based analytical model to incorporate both the prior sentiment information and newly collected sentiment observations; (2) reviewing various public sentiment measures in terms of their formulas, boundaries, and applications; and (3) deriving a systematic approach to include uncertainty of public sentiment reliably. For practitioners, this proposed approach is capable of (1) selecting a public sentiment measure based on the corresponding application; and (2) ranking public sentiment in a more reliable way by including uncertainty for planning actions.

Although this research proposes an approach for incorporating uncertainty of public sentiment caused by the statistical variation of the sampled dataset, the uncertainty caused by the disagreements among multiple sentiment classifiers is not included. In the future, the authors will incorporate such uncertainty with the statistical variation for a more advanced public sentiment modeling.

**ACKNOWLEDGMENTS**

This study is supported by the Thomas F. and Kate Miller Jeffress Memorial Trust and the National Science Foundation (Grant No. 1761950). Any opinions, findings, and conclusions or recommendations expressed in this material are those of the authors and do not necessarily reflect the views of the Thomas F. and Kate Miller Jeffress Memorial Trust and the National Science Foundation. The authors also gratefully acknowledge the support of internal grants from the Global Resilience Institute, Northeastern University.

## AUTHOR BIOGRAPHIES

**YUDI CHEN** is a PhD student in the Department of Civil, Environmental & Infrastructure Engineering, George Mason University. His research focuses on the integration of data mining and complex system simulation to enhance infrastructure and social resilience. His email address is ychen55@masonlive.gmu.edu.

**WENYING JI** is an Assistant Professor in the Department of Civil, Environmental & Infrastructure Engineering, George Mason University. He received his PhD in Construction Engineering and Management from the University of Alberta. He is an interdisciplinary scholar focused on the integration of advanced data analytics, complex system simulation, and construction management to enhance the overall performance of infrastructure systems. His e-mail address is wji2@gmu.edu.

**QI WANG** is an Assistant Professor in the Department of Civil and Environmental Engineering, Northeastern University. He received a PhD from the Department of Civil and Environmental Engineering at Virginia Tech. His research focuses on the interplay between urban information and mobility, infrastructure and social resilience. His email address is ryan.qi.wang@gmail.com.